\documentclass[pra,twocolumn,showpacs]{revtex4}

\usepackage[english]{babel}
\usepackage{bm}
\usepackage{amsmath}
\usepackage{amssymb}
\usepackage{graphicx}
\usepackage[normal]{subfigure}

\begin{document}

\title{Ferroplasmas: Magnetic Dust Dynamics in a Conducting Fluid}
\author{J. Zamanian, G. Brodin, and M. Marklund}
\affiliation{Department of Physics, Ume{\aa} University, SE--901 87 Ume{\aa}, Sweden}

\begin{abstract}
We consider a dusty plasma where the dust particles have a magnetic dipole moment. A Hall-MHD type of model, generalized to account for the intrinsic magnetization, is derived. The model is shown to be energy conserving, and the energy density and flux is derived.  
The general dispersion relation is then derived, and we show that kinetic Alfv\'{e}n waves exhibit an instability for a low temperature and high density plasma. We discuss the implication of our results.
\end{abstract}

\maketitle

\section{Introduction}
Plasmas with impurity particles, so called dusty or complex plasmas, have a manifold of applications, ranging from applied to basic research problems. Due to this wide range of applicability, there has been a steadily increasing interest in the physics of dusty plasmas (see Ref.\ \cite{shukla} and references therein). The above mentioned spectrum of applications covers, e.g., planetary rings, lightning discharges in smoke contaminated air, fusion plasmas, low-temperature laboratory plasmas, and processing plasmas in the semi-conductor industry. To be specific, by a dusty plasma one usually means a three component plasma consisting of electrons, ions, and dust which is considered to be significantly heavier than the ions \cite{pandey}. The charge of the dust is assumed to range from a few electron charges to thousands. For astronomical applications it is often important to also include the dynamics of neutral particles \cite{birk}. Dusty plasmas contains novel physical phenomena, such as dust acoustic waves \cite{rao}, dust ion-acoustic waves \cite{shukla92}, dusty plasma crystals \cite{chu,thomas}, and dust lattice waves \cite{melandso}, all of which have been experimentally verified, see Refs.\ \cite{merlino,shukla} and references therein. In fact, due to the (in general) relatively low phase velocity of dusty plasma waves, these plasmas are useful for probing basic properties of plasma excitations. 

Lately there has also been an increasing interest in quantum plasma physics \cite{gardner,manfredi,haas,haas00,garcia,kuzmenkov,brodin07}, in particular the nonlinear aspects of such systems. Such plasmas are in general typical for condensed matter environments, where the density of the electron gas is high \cite{manfredi,Pines}, giving a considerable influence from the wave function structure of the electrons. Moreover, the spin properties of plasmas has been investigated recently by means of quantum hydrodynamical models \cite{brodin07,marklund07,marklund071} and also by the use of spin kinetic models \cite{cowley,kulsrud,newarticle}. Even in a regime considered as classical, the effects of spin may give a nontrivial influence on the dynamics of an electron plasma \cite{brodin-marklund-manfredi}. Furthermore, spin and the intrinsic magnetic moment of the constituent particles are an essential part of magnetic fluids or ferrofluids \cite{rosensweig}. A ferrofluid is a mixture of nanosized magnetic particles suspended in a liquid. In this article we will consider a model where magnetic dust particles are suspended in an electron-ion plasma, which can be said to be the plasma analogue of a ferrofluid. Such systems has recently been investigated both theoretically, considering single particle dynamics  \cite{malnev,malnev2,uchida,tsytovich03,vladimirov} and experimentally \cite{yaroshenko03,samsonov}. Here we will consider sufficiently low frequency phenomena so that we may use a hydrodynamical model to describe the dynamics of the dust particles.

In Section \ref{model} we present the governing equation and show that the model satisfies an energy conservation law in which magnetization transport is included. We then linearize the equations in Section \ref{linear} to obtain the general dispersion relation. For the case of a static external magnetic field we consider modes propagating perpendicular to the magnetic field as well as kinetic Alfv\'en waves. We show that these modes exhibit instabilities. Finally, in Section \ref{conclusions} we summarize and draw our conclusions.

\section{A magnetized dust model} \label{model}

We here consider a three component plasma consisting of electrons, positive ions and negatively charged dust particles (denoted by subscripts $e$, $i$ and $d$ respectively). The dust is assumed to be magnetized and have a charge $-Z_d e$ (where $e$ is the elementary charge). The magnetization of the dust can be assumed to be due to quantum mechanical spin, or from a macroscopic magnetization of the dust grains themselves. The size of the dust particles in, e.g., astronomical environment ranges from a few nanometers to about 100 microns and the weight from about $10^{-15}$g to $10^{-5}$g \cite{pandey,shukla}. In the laboratory for example Ref.\ \cite{samsonov} uses grains with size $4.5 \mu$m and mass $\sim 10^{-14}$g. Hence the dust is much heavier than the ions, $m_d \gg m_i$. 

In the framework of the multi-fluid theory the dynamics is governed by the continuity and momentum conservation equations. The continuity equation is given by
\begin{equation}
	\frac{\partial n_s}{\partial t} + \nabla \cdot (n_s \mathbf v_s) = 0 ,
\end{equation}
where the subscript $s$ denotes the different species ($s=e,i,d$), $n_s$ is the number density and $\mathbf v_s$ is the fluid velocity of particle type $s$. The momentum conservation for the dust reads
\begin{eqnarray}
	&&
	m_d n_d \left( \frac{\partial}{\partial t} + \mathbf v_d \cdot \nabla \right) \mathbf v_d = 
	- Z_d e n_d (\mathbf E 	+ \mathbf v_d \times \mathbf B) 
	\nonumber \\ &&\qquad\qquad\qquad\qquad\qquad
	- k_B \nabla ( T_d n_d) + M_\alpha \nabla B_\alpha ,
	\label{2}
\end{eqnarray}
where $n_d$ is the number density of the dust, $k_B$ is Boltzmann's constant, $T_d$ is the dust temperature and $\mathbf M$ is the magnetization. We use Einstein's summation convention for greek indices. 
The last term in the equation above is usually neglected when considering plasmas, but for microsized dust grains in a plasma the mutual magnetic dipole interaction can be of importance for laboratory conditions \cite{yaroshenko03}. 
We will consider perturbations slow compared to the plasma frequencies of the ions and the electrons. Neglecting the momentum of these particles we obtain
\begin{eqnarray}
	0 &=& - e n_e ( \mathbf E + \mathbf v_e \times \mathbf B ) - k_B \nabla (T_e n_e) 
	\label{3}
	\\
	0 &=& Z_i e n_i ( \mathbf E + \mathbf v_i \times \mathbf B ) - k_B \nabla (T_i n_i) ,
	\label{4}
\end{eqnarray}
where $T_{e(i)}$ is the temperature of the electrons (ions) and $Z_i e$ is the charge of the ions. Note, we neglect the spin of the electron and the protons. The magnetization of the dust is assumed to be orders of magnitude larger. The equations above are coupled to Maxwell's equations
\begin{eqnarray}
	\nabla \cdot \mathbf E &=& - \frac{e}	{\epsilon_0}
	( n_e - Z_i n_i + Z_d n_d) , 
	\label{5}\\
	\nabla \times \mathbf E &=& - \frac{\partial\mathbf B}{\partial t}
	\label{6}, 
\end{eqnarray}
and
\begin{equation}
	\nabla \times \mathbf B = \mu_0 \nabla \times \mathbf M - e \mu_0
	(n_e \mathbf v_e - Z_i n_i \mathbf v_i + Z_d n_d \mathbf v_d) .
	 \label{7}
\end{equation}
A closed system of equations for the dust can be derived. To do this we start by adding the momentum Eqs.\ \eqref{3} and \eqref{4} and assume that quasi-neutrality holds $Z_i n_i \approx n_e + Z_d n_d$ to obtain
\begin{equation}
	-e Z_d n_d \mathbf E = - e(n_e \mathbf v_e - Z_i n_i \mathbf v_i) \times 
	\mathbf B - k_B \nabla (T_i n_i + T_e n_e) \label{8} .
\end{equation}
Solving this for the electric field and inserting it into the momentum equation for the dust \eqref{2} yields
\begin{eqnarray}
	&& m_d n_d \left(
	\frac{\partial}{\partial t} + \mathbf v_d \cdot \nabla\right)
	\mathbf v_d 
	= 
	\nonumber \\
	&& \quad \quad \quad
	-e (n_e \mathbf v_e - Z_i n_i \mathbf v_i + Z_d n_d \mathbf v_d ) 
	\times \mathbf B 
	\nonumber \\ 
	&& \quad \quad \quad
	- k_B \nabla (T_d n_d + T_i n_i + T_e n_e) + M_\alpha \nabla B_\alpha .
\end{eqnarray}
Using Eq.\ \eqref{7} we obtain
\begin{eqnarray}
	&&
	m_d n_d \left(
	\frac{\partial}{\partial t} + \mathbf v_d \cdot \nabla\right)\mathbf v_d 
	= 
	\nabla \times \left(\frac{\mathbf B}{\mu_0} - 
	\mathbf M \right) \times 	\mathbf B 
	\\ && 
	- k_B \nabla \left[\left(T_e + \frac{T_i}{Z_i}\right)n_e 
	+ \left(T_d \!+ \! \frac{Z_d T_i}{Z_i} \right)n_d \right] 
	+ M_\alpha \nabla B_\alpha \nonumber
\end{eqnarray}
where the quasi-neutrality condition has been used to rewrite the thermal pressure terms and we will assume that $(T_e + T_i/Z_i)n_e \ll (T_d + Z_d T_i/Z_i)n_d$ in order to obtain a closed set of equations. This condition can of course be relaxed, introducing new thermal effects. However, the principle dynamics of the complex plasma will not be significantly affected. Using some vector identities we can finally write the continuity and momentum equations for the dust 
\begin{subequations} \label{system}
\begin{equation}
	\frac{\partial n_d}{\partial t} + \nabla( n_d \mathbf v_d ) = 0 , 
\end{equation}
and 
\begin{eqnarray}
	&&
	m_d n_d \left(
	\frac{\partial}{\partial t} + \mathbf v_d \cdot \nabla\right) \mathbf v_d 
	= \mathbf B \cdot \nabla \left( \frac{\mathbf B}{\mu_0} - \mathbf M \right)
	\nonumber \\ && \quad
	- \nabla \left[ \frac{B^2}{2\mu_0} \! - \!
	\mathbf M \cdot \mathbf B \! + \!
	k_B \left( T_d \! + \! \frac{Z_d T_i}{Z_i} \right) n_d \right].
\end{eqnarray}
respectively. Using Eqs.\ \eqref{6}, \eqref{7} and \eqref{8} we can derive the time evolution equation for the magnetic field
\begin{equation} 
	\frac{\partial \mathbf B}{\partial t} 
	= \nabla \times (\mathbf v_d \times \mathbf B) + 
	\nabla \times \left\{ \frac{\left[ \nabla \times ( \mathbf B - \mu_0 \mathbf M) \right] 
	\times \mathbf B }{\mu_0 Z_d e n_d} \right\} ,
\end{equation}
The magnetization is taken to be proportional to the density of the dust particles and in the direction of the magnetic field
\begin{equation}
	\mathbf M = \mu_d ( B , T_d ) n_d \hat{ \mathbf B } .
	\label{11d}
\end{equation}
\end{subequations}
This model includes the case where the particles have an intrinsic magnetic moment that will be aligned with an applied magnetic field, which is what we have in mind. However, it also includes the case where the dust particles have no net magnetic moment. The occurrence of an external magnetic field will induce a magnetization in the dust particles which yields a macroscopic magnetization of the fluid. A magnetization of the dust could also arise from spinning of the dust particles (with charges attached to the surface) which yields a diamagnetic response to an applied field. However, the magnetic dust-dust interaction due to this can often be neglected. See Ref.\ \cite{tsytovich03} for a more detailed discussion.
Our model can be compared with a ferrofluid which is a colloidal suspension of magnetic particles in a liquid. In an \textit{ionic} ferrofluid the magnetic particles are kept apart by repulsive electrostatic forces, see e.g. Refs.\ \cite{rosensweig,bringuier,wang,szalai,samsonov}. Due to the high density and the correspondingly high collision frequency, we note that free currents can typically be neglected in ionic ferrofluids. By contrast, the simultaneous existence of free currents and magnetic dipole moment have been shown to be significant in dusty plasmas \cite{samsonov}. A more general assumption than Eq.\ \eqref{11d} would be to assume that the magnetization is 
also dependent on the electron and ion temperatures since collisions with 
these particles may change the magnetization. 

With the exception for the occurrence of the 
magnetization due to the magnetic moment of the dust, $\mathbf M$, these 
equations are the same as Hall-MHD theory \cite{lighthill,witalis}. 
Moreover, it should be noted that the structure
of the system of Eqs.\ (11a)-(11c) is the same as one get from the magnetized
ideal MHD model of Refs. \cite{brodin07,brodin06}, if that system is extended to include the Hall current. In that case naturally the dust
density and velocity instead will refer to the ion density and
velocity. Although Eqs.\ (11a)-(11c) \emph{can} describe an electron-ion plasma, it
should be noted that the physics for that case is different in several 
respects: Firstly,
in the case of an electron-ion plasma, it is the lighter species, the
electrons, that contribute to the magnetization, due to their magnetic
moments being larger than those of the ions. Secondly we stress that the
validity conditions of the model have no simple correspondence between the
dust dominated plasma and the electron-ion plasma case. In what follows, we
will mainly be concerned with the dusty plasma applications.

Next we want to find an energy conservation law for the Eqs.\ \eqref{system}.
In order to do this we must specify an equation of state for the system. For simplicity we choose the simple model 
\begin{equation}
	\frac{P}{P_0} = \left( \frac{n_d}{n_{d0}} \right)^{\gamma}, 
\end{equation}
for the \textit{total pressure} $P=k_B(T_d + Z_d T_i/Z_i)$, 
where $P_0$ and $n_{d0}$ are the equilibrium pressure and density respectively. With this equation of state the energy conservation becomes
\begin{equation}
	\frac{\partial W}{\partial t} + \nabla \cdot \mathbf P = 0 , 
\end{equation}
where 
\begin{equation} \label{energy}
	W = \frac{m_d n_d v_d^2}{2} + \frac{P}{\gamma -1} + 
	\frac{B^2}{2\mu_0} - \mathbf B \cdot \mathbf M
\end{equation}
is the energy per volume and 
\begin{eqnarray}\label{flow}
	\mathbf P &=& \frac{m_d n_d v_d^2}{2} \mathbf v 
	+ \frac{\gamma P}{\gamma - 1}  \mathbf v	
	- (\mathbf B \cdot \mathbf M)  \mathbf v 
	\nonumber \\
	&&
	- \left[\mathbf v \times \mathbf B + \frac{1}{Z_d e n_d} 
	(\nabla \times \mathbf H) \times \mathbf B \right] \times \mathbf H 
\end{eqnarray}
is the flow of energy out of the region. In Eq.\ \eqref{energy} the first term is the kinetic energy per volume, the second term is the energy density from the pressure, the third term is the energy stored in the magnetic field and the last term is the energy in each volume element due to the magnetic moment of the dust particles. Similarly, in Eq.\ \eqref{flow}, the first three terms are the flow of kinetic, pressure and magnetic energy density that follows the flow of each volume element. The last term is the Poynting vector which is modified by the inclusion of the Hall-term and the magnetization.

\section{The dispersion relation} \label{linear}
We linearize Eqs.\ \eqref{system} and Fourier decompose. Furthermore, the coordinate system is defined so that $\mathbf B_0 = B_0 \hat{\mathbf z}$ and $\mathbf k = k_x \hat{\mathbf x} + k_z \hat{\mathbf z}$. For simplicity we choose an isothermal pressure model. This gives the dispersion relation
\begin{widetext}
\begin{eqnarray}
	\left\|
		\begin{array}{ccc}
			\omega^2 - k_x^2 (\tilde V_{da}^2 - V_{dB}^2) - k^2 \tilde V_{dA}^2
			&
			\displaystyle{- i\frac{\omega}{\omega_{cd}} ( k^2 \tilde V_{dA}^2 - k_x^2 V_{dB}^2 )}
			&
			- k_x k_z \tilde V_{da}^2
			\\[3mm]
			\displaystyle{i\frac{\omega}{\omega_{cd}} k_z^2 \tilde V_{dA}^2}
			&
			\omega^2 - k_z^2 \tilde V_{dA}^2
			&
			\displaystyle{- i\frac{\omega}{\omega_{cd}} k_x k_z \tilde V_{dA}^2}
			\\[3mm]
			- k_x k_z \tilde V_{da}^2
			&
			\displaystyle{i\frac{\omega}{\omega_{cd}} k_x k_z V_{dM}^2}
			&
			\omega^2 - k_z^2 V_{da}^2
		\end{array}
	\right\|
	= 0 ,
	\label{disprel}
\end{eqnarray}	
\end{widetext} 
where $\tilde V_{dA}^2 \equiv V_{dA}^2 - V_{dM}^2$, $\tilde V_{da}^2 \equiv V_{da}^2 - V_{dM}^2$ and we have defined
\begin{subequations}\label{vel}
\begin{eqnarray}
	V_{da}^2 &=& \frac{k_B}{m_d} \left( T_d + \frac{Z_d T_i}{Z_i} \right)	\\
	V_{dA}^2 &=& \frac{B_0^2}{\mu_0 m_d n_{d0}} \\
	V_{dM}^2 &=& \frac{\mu_{d0} B_0}{m_d} \\
	V_{dB}^2 &=& \frac{\partial \mu_{d0}}{\partial B_0} \frac{B_0^2}{m_d} .
\end{eqnarray}
\end{subequations}
The velocity $V_{da}$ is a generalized thermal speed for the dust, $V_{dA}$ is the dust Alfv\'en speed and $V_{dM}$ and $V_{dB}$ are related to the magnetization of the dust.  
The frequency, $\omega_{cd} = Z_d e B_0/m_d$
is the cyclotron frequency of the dust. The dispersion relation, Eq.\ \eqref{disprel}, is in general a third degree polynomial in $\omega^2$. Specifically, for $\omega \ll \omega_{cd}$ the three roots to the dispersion relation are the fast and slow magnetosonic modes, and the shear Alfv\'en wave \cite{Brodin-Marklund-PRE}.

To see the implications of the derived dispersion relation a couple of special cases are now considered. For a wave propagating perpendicular to the magnetic field ($\mathbf k = k_x \hat{\mathbf x}$) the dispersion relation is obtained from Eq.\ \eqref{disprel} and reads
\begin{equation}
	\omega^2 = k_x^2 \left[ \tilde V_{dA}^2 + \tilde V_{da}^2 - V_{dB}^2 
	\right] . 
	\label{14}
\end{equation}
To get a qualitative description of the instability condition for Eq.\ \eqref{14} we assume that the dust and the ions are in thermal equilibrium so that we may assume $(T_d + Z_d T_i/Z_i) = N T_d$ where $N>1$ is a constant. Typically we have $Z_d/Z_i$ ranging from unity to a few thousands \cite{shukla}. Furthermore we assume that the spins are thermally distributed (see e.g. Ref.\ \cite{Brodin-Marklund-PRE}) the magnetization per density can be written 
\begin{equation}\label{eq:magmom}
	\mu_{d} = \bar\mu_d \tanh \left( \frac{\bar \mu_d B}{k_B T_d} \right) .
\end{equation}
In the case that the dust particles have a high total spin number the $\tanh$-function in Eq.\ \eqref{eq:magmom} corresponding to Fermi-Dirac statistics should be replaced by the 
Langevin-function, corresponding to Maxwell-Boltzmann statistics. The 
difference between these functions are relatively small, however,
and thus we will use Eq.\ \eqref{eq:magmom} for the remainder of this paper. We note that the magnetic moment $\bar\mu_d$ can be several orders of magnitude larger than the Bohr magneton. The condition for instability, Eq.\ \eqref{14}, then becomes
\begin{equation}
 	\frac{B_0}{\mu_0 n_{d0} \bar \mu_d} 
 	+ N \frac{k_B T_d}{\bar \mu_d B_0}
 	- 2 \frac{\mu_{d0}}
 	{\bar\mu_d}  - \frac{\bar\mu_d B_0}{k_B T_d} \left[ 1 - 
 	\left( \frac{\mu_{d0}} {\bar\mu_d} \right)^2 \right] < 0 ,
 	\label{16} 
\end{equation}
where $\mu_{d0} \equiv \mu_d(B_0)$ in accordance with Eq.\ (\ref{eq:magmom}). Note that this equation implies that more of the magnetic dipoles are 
re-oriented towards the lower energy state
when the magnetic field strength is increased. For this to apply, the 
relaxation time to reach the lower energy state must be shorter than the 
wave period time. In case the opposite ordering holds, the fraction of 
particles in the different energy states remain constant during a wave 
period, and consequently the term proportional to $V_{dB}^2$ in Eqs.\ \eqref{disprel} and \eqref{14} should be dropped, which correspond to neglecting the fourth and  
last term in Eq.\ \eqref{16}. 
The difference in the dispersion relation, depending 
on whether the magnetic dipoles have time to change during a wave period 
or not, turns out to be relatively small, however. For definiteness, we 
will stick to the case where the relaxation time of the magnetic dipoles 
is sufficiently fast for Eq.\ \eqref{16} to apply for the rest of this work.

Next we consider the kinetic Alfv\'en type of waves. In this case the ordering $k_z \ll k_x$, $V_{da} \ll V_{dA}$ and $\omega \sim k_z V_{dA}$ applies. The dispersion relation can then be approximated by
\begin{eqnarray}\label{Eq:21}
	&& \frac{\omega^2}{k_z^2 \tilde V_{dA}^2} = 
	\\
	&&
	1 \! + \! \frac{\omega^2}{\omega_{cd}^2} 
	\frac{k_x^2\left[(\tilde V_{dA}^2 \! - \! V_{dB}^2)\tilde V_{da}^2  
	+ V_{dM}^2(\tilde V_{dA}^2 \! + \!  \tilde V_{da}^2 \! - 
	\! V_{dB}^2)\right]}
	{(\tilde V_{dA}^2 + \tilde V_{da}^2 - V_{dB}^2)
	(\omega^2 - k_z^2 V_{da}^2) + k_z^2 \tilde V_{da}^4} 
	\nonumber
\end{eqnarray}
For $V_{dM}^2, V_{dB}^2 \rightarrow 0$ this reduces to the well known kinetic Alfv\'en waves \cite{hasegawa82}. Analyzing Eq.\ \eqref{Eq:21} it is seen that the wave-mode can be unstable provided the numerator of the second term of the right hand side is negative, i.e. we obtain the instability condition
\begin{equation}
	(\tilde V_{dA}^2 - V_{dB}^2)\tilde V_{da}^2  
	+ V_{dM}^2(\tilde V_{dA}^2 + \tilde V_{da}^2 - V_{dB}^2) < 0 
	\label{18}
\end{equation}
Assuming once more that the dust and the ions are in thermal equilibrium we get 
\begin{equation}
	\frac{B_0}{\mu_0 n_{d0} \bar \mu_d} 
	- \frac{\bar \mu_d B_0}{k_B T_d} 
	+ \frac{(N-1)}{N} \frac{\bar \mu_d B_0}{k_B T_d} 
	\tanh^2 \left( \frac{\bar \mu_d B_0}{k_B T_d}  \right)	
	< 0. 
	\label{kininst}
\end{equation} 

\begin{figure*}
\subfigure[]{\includegraphics[width=0.45\textwidth]{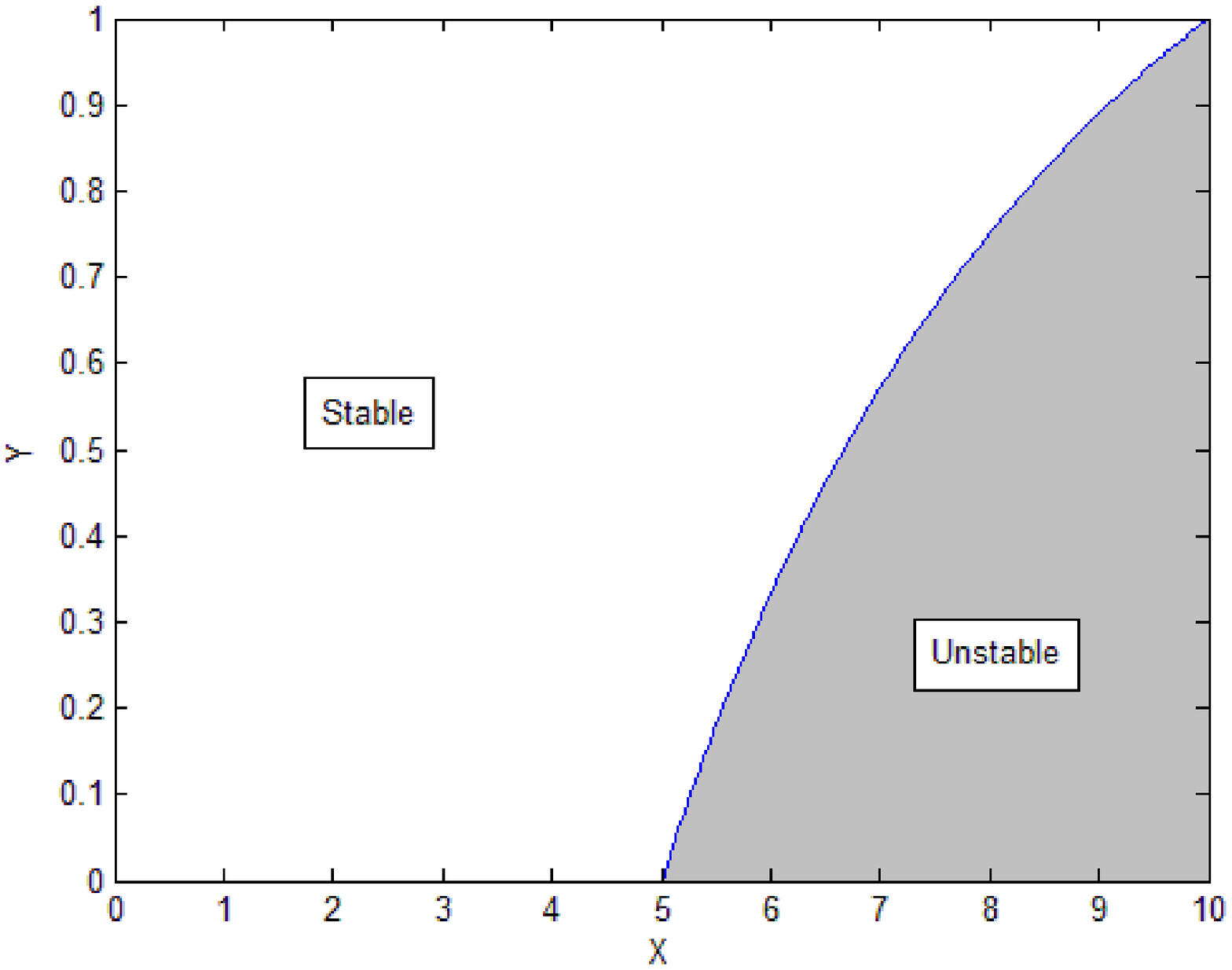}}
\subfigure[]{\includegraphics[width=0.45\textwidth]{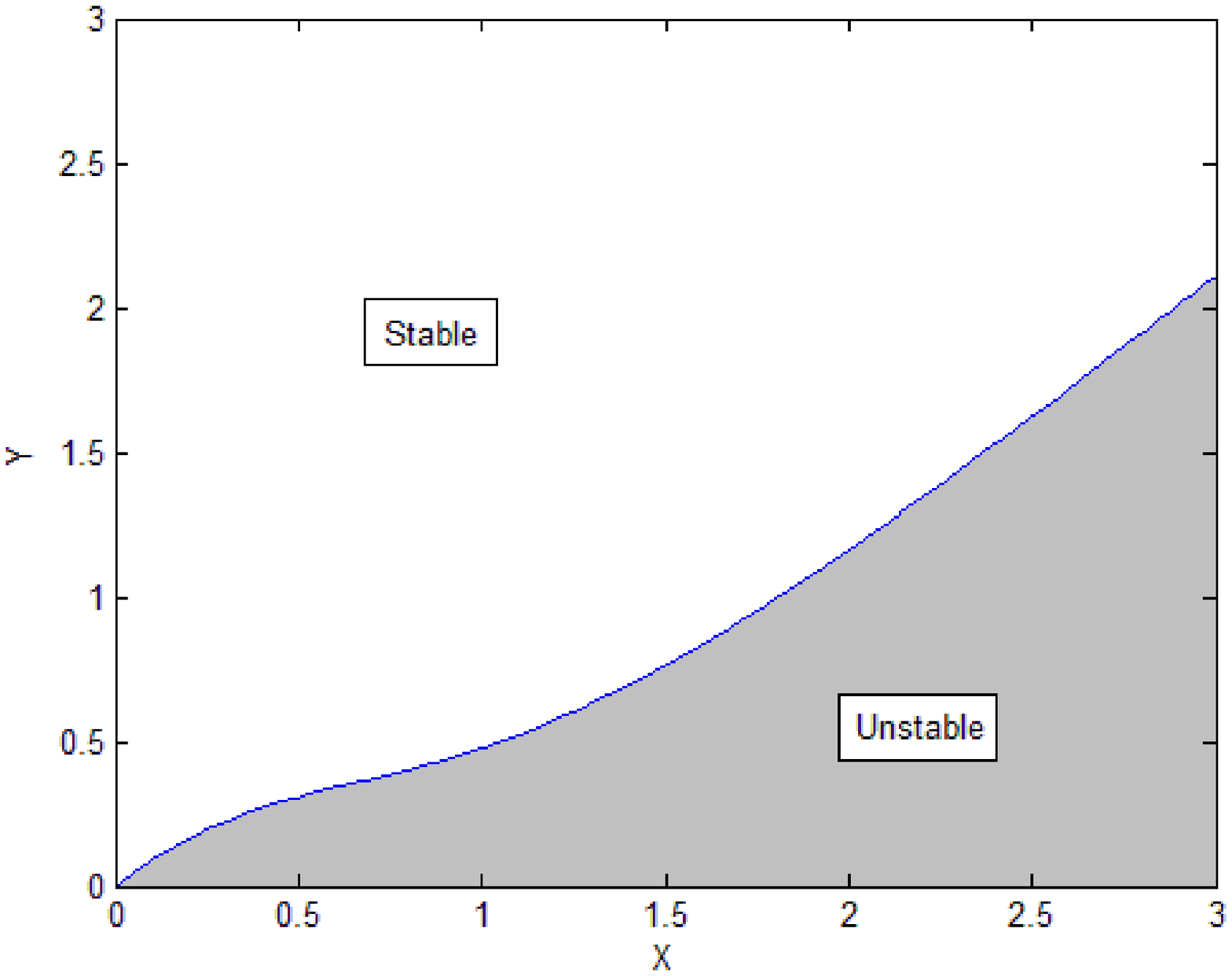}}
	\caption{The instability regions for the (a) the perpendicular mode, Eq.\ \eqref{16} and the (b) kinetic Alfv\'en mode, Eq.\ \eqref{kininst} plotted in terms of the dimensionless parameters $X=\bar\mu_dB_0/(k_BT_d)$ and $Y=B_0/(\mu_0n_{d0}\bar\mu_d)$ with $N=10$.}
	\label{fig:instability}
\end{figure*}

The conditions for instabilities Eqs.\ \eqref{16} and \eqref{kininst} have been plotted in Fig.\ \ref{fig:instability}. Note that in order to have instabilities we need to have sufficiently high densities and/or sufficiently low temperatures. 

We here give the following simplified picture of why the instability of this type can occur. The volume elements of the plasma are electrically neutral since the electron and ion background will screen any excess electrical charge. Further, the magnetization of a volume element is in the direction of the magnetic field and the different volume elements will attract each other like small magnets. Consider now the magnetic flux through a surface with normal parallel to $\mathbf B$. If the oscillations have $k_x=0$ then the magnetic flux through the surface will not change, and hence there will be no build up of magnetization. If, on the other hand, the oscillations occur perpendicular to the field there can be a local build up of the magnetic field. This can for sufficiently low temperature and high density cause the plasma to collapse similarly to the case of the Jeans instability \cite{Brodin-Marklund-PRE,Herrera-Santos}.

\section{Summary and Conclusion}\label{conclusions}
In the present paper we have put forward a Hall MHD type
of model with an intrinsic magnetization. We have shown that the 
magnetized Hall MHD model is energy conserving, and presented the 
expressions for energy density and the energy flux.
A set of equations of this type could describe different types of systems: 
An ordinary electron-ion plasma, in which case the magnetization would be
due to the electron spin, or - as emphasized here - a three component 
plasma containing electrons, ions and heavy dust particles.

As the next step, we have investigated the linear modes and the stability
properties of the homogeneous system. The general dispersion relation has
been derived, describing the fast and slow magnetosonic modes and the shear
Alfv\'{e}n waves, as modified by the Hall current and the magnetization of
the system. Due to the magnetization, the homogeneous system may be
unstable, as predicted already from an magnetized ideal MHD type of model
\cite{Brodin-Marklund-PRE}. The main new finding from the stability 
analysis in this paper, is that inclusion of the effects due to the Hall 
current extends the unstable region of parameter space, as described by 
Fig 1. The wave that first becomes unstable turns out to be the magnetized 
version of the kinetic Alfv\'{e}n wave. 

For the effects of magnetization to be significant, we need relatively high
density plasmas, and/or low temperatures. For the case of electron-ion
plasmas, these can be found in astrophysical environments, such as the
interior of white dwarf stars and pulsars. 

We can make some numerical estimations for the parameters $X=\bar\mu_dB_0/(k_BT_d)$ and $Y=B_0/(\mu_0n_{d0}\bar\mu_d)$. From Ref.\ \cite{yaroshenko03} we find that a the magnetic moment per particle can be of the order $10^{-12} \textrm{m}^{-2} \textrm{A}^{-1}$ and magnetic induction of the order $0.1\textrm{T}$. Furthermore, we assume that the dust temperature is low $T_d \sim 1 \textrm{K}$. The density of particles is taken to be $n_{d0} \sim 10^{12}\textrm{m}^{-3}$ as in Ref.\ 
\cite{malnev2}. 
This gives us $X\approx 10^{10}$ and $Y \approx 10^6$. Comparing this with 
Figs.\ \ref{fig:instability} we see that the instabilities considered here 
are not possible to detect in current experiments. 
As shown e.g. by Refs.\ \cite{yaroshenko03,samsonov}, however, magnetic dipole effects can still be of significance in dusty plasmas. 

The model developed here should be considered only as a first step since it does not account for potentially important effects of more elaborate models. 
These include the two-fluid model, where spin up and spin down populations are considered as different species \cite{brodin-marklund-manfredi}, the kinetic description \cite{newarticle} and models including nearest neighbor interactions \cite{vladimirov,yaroshenko03}, which is important in the strongly coupled regime.

\begin{acknowledgments}
This research was supported by the European Research Council under  
Contract No.\ 204059-QPQV, and the Swedish Research Council under  
Contract No.\ 2007-4422.
\end{acknowledgments}


\end{document}